\documentclass[11pt,a4paper]{emulateapj}
\usepackage{graphicx, latexsym, booktabs, multirow, color}
\usepackage{amsmath, footnote, epsfig, natbib}
\pdfoutput=1

\def\beq{\begin{equation}}
\def\eeq{\end{equation}}
\def\bey{\begin{eqnarray}}
\def\eey{\end{eqnarray}}
\def\lsim{\mathrel{\raise.3ex\hbox{$<$\kern-.75em\lower1ex\hbox{$\sim$}}}}
\def\gsim{\mathrel{\raise.3ex\hbox{$>$\kern-.75em\lower1ex\hbox{$\sim$}}}}

\begin{document}

\title{ High-Energy Neutrinos from Sources in  Clusters of Galaxies}
\author{Ke Fang\altaffilmark{1,2}, Angela V. Olinto \altaffilmark{3}}
\altaffiltext{1}{University of Maryland, Department of Astronomy, 1105 PSC, College Park, MD 20742, USA}
\altaffiltext{2}{Joint Space-Science Institute, College Park, MD, 20742}
\altaffiltext{3}{Department of Astronomy \& Astrophysics, Kavli Institute for Cosmological Physics, The
  University of Chicago, Chicago, Illinois 60637, USA}

\begin{abstract}
High-energy cosmic rays can be accelerated  in clusters of galaxies,  by mega-parsec scale shocks induced by accretion of gas during the formation of large-scale structure, or by powerful sources harbored in clusters. Once accelerated, the highest energy particles leave the cluster via almost rectilinear trajectories, while lower energy ones can be confined by the  cluster magnetic field up to cosmological time and interact with the intracluster gas. Using a realistic model of the baryon distribution and the turbulent magnetic field in clusters, we   studied the propagation and hadronic interaction of high-energy protons in the intracluster medium.  We report the cumulative cosmic ray and neutrino spectra generated by galaxy clusters including embedded sources,  and demonstrate that   clusters can  contribute a significant fraction of the observed IceCube neutrinos above 30 TeV while remaining undetected in high-energy cosmic rays and $\gamma$ rays for reasonable choices of parameters and source scenarios.

\end{abstract}
\maketitle

\section{Introduction}
\label{sec:introduction}

%IceCube observation - suggested spectra index
The IceCube  Observatory has observed the first  high-energy astrophysical neutrinos \citep{Aartsen:2013jdh, Aartsen:2013bka, Aartsen:2013uuv}.  The all-flavor diffuse neutrino flux is reported to be $\Phi_\nu=2.06\times10^{-18}\,(E_\nu/10^5\,\rm GeV)^{-2.46}\,\rm GeV^{-1}cm^{-2}sr^{-1}s^{-1}$ for the energy range $25\,\rm TeV < E_\nu < 1.4\,\rm PeV$ \citep{2015PhRvD..91b2001A}. 
%The arrival directions indicate that most of the observed neutrinos  come from extragalactic sources \citep{Ahlers:2015moa}.
Searches for small-scale anisotropies from the three-year IceCube data do not see any significant clustering or correlations \citep{Aartsen:2014ivk}.  The origin of these high-energy neutrinos remains unclear \citep{MuraseReview}.

Accretion of gas during the formation of the large-scale structure  can give rise to mega-parsec scale shocks that accelerate high-energy cosmic rays \citep{2000ApJ...542..608M, Ryu:2003cd}, and even ultra-high energy cosmic rays (UHECRs) if the medium around the shocks contain heavy nuclei \citep{Inoue07}. Numerical simulations \citep{Vazza:2014usa,  2014ApJ...785..133H, 2015ApJ...800...60M} also suggest the possibility of stochastic particle acceleration  in large-scale structures, though the flux level of cosmic rays  depend on the Mach number and the location of shocks \citep{Vazza:2014usa}.

Powerful sources harbored in the galaxy clusters can also accelerate particles to high energies. Many plausible candidate sources have been proposed in literature, including steady sources like active galactic nuclei (AGN) (e.g. \cite{PhysRevLett.66.2697, Winter:2013cla, Murase:2014foa, Dermer:2014vaa}), and 
  transients like gamma-ray bursts (GRBs)  (see \cite{0034-4885-69-8-R01} for review), fast-spinning newborn pulsars \citep{Blasi00, Fang:2012rx, Fang:2013vla}, magnetars \citep{Arons:2002yj, 2009PhRvD..79j3001M}, and blazar flares \citep{Farrar:2008ex}.

Clusters are famously  known as   cosmic ray reservoirs, due to their ability to confine cosmic rays with turbulent magnetic fields up to cosmological time (\citet{1996SSRv...75..279V, 0004-637X-487-2-529},  also see   \cite{2015ASSL..407..557B} for a recent review). Hence the accelerated cosmic rays  have a good chance to interaction with the intracluster medium (ICM), leading to the production of secondary neutrinos and $\gamma$ rays. 
 A cosmic ray reservoir scenario is favored to explain the absence of detection of neutrinos above a few PeV, especially around the Glashow resonance at 6.3 PeV. This is because rather than otherwise a  mysterious stop in the cosmic ray spectrum,  a    ``cosmic ray reservoir" scenario naturally introduces a spectral softening   caused by the faster escape of the higher-energy cosmic rays in a magnetized source environment.

Many analytical and semi-analytical works were conducted to calculate these secondary fluxes   \citep{1995PhRvL..75.3052D, 1996SSRv...75..279V, 0004-637X-487-2-529,  2006PhRvD..73d3004D, Murase:2008yt, 2008ApJ...687..193W, Murase:2013rfa}. However most of the analytical approaches adopted an overly simplified modeling of the  ICM by assuming both uniform gas distribution and uniform magnetic field. \cite{1998APh.....9..227C} took into account the density profile and mass function of galaxy clusters, yet  still assumed a uniform magnetic field within each   cluster and the same cosmic ray diffusion length  for all groups of clusters with different masses.  Numerical propagation of cosmic rays in more realistic three-dimensional cluster magnetic fields were explored  by \cite{2004APh....22..167R} and \cite{Kotera09}, but the studies were limited to neutrinos from a single type of clusters.  

 Numerical modeling of the non-uniform gas distribution and magnetic field structure is crucial  in an environment like the ICM, because both  duration and  location of the confinement  of the charged particles impact the interaction rate,  and thus an accurate computation of the neutrino production.  Moreover, 
 although massive clusters could be brighter cosmic ray and neutrino sources, they are far less common than medium-sized clusters. It  is  thus unclear how the mass and redshift dependence of cluster number density would impact the total neutrino spectrum. 
It is also unknown which group of clusters
would  dominate PeV cosmic rays and thus  be more relevant for the detected neutrinos. The latter point is also important for revealing  the possibility of pinpointing  sources with the increasing statistics  of IceCube as well as future experiments  \citep{Ahlers:2014ioa, Fang:2016hyv}.

\cite{Bechtol:2015uqb} pointed out that new studies of the blazar flux at gamma rays above 50 GeV result in a lower residual non-blazar component of the extragalactic gamma-ray background. This puts a tight constraint on   neutrino sources that are transparent to gamma rays \citep{Murase:2015xka}, 
especially those which produce neutrinos through hadronuclear (pp) interactions (as in galaxy clusters  \citep{Kotera09}). However, this constraint is drawn based on the assumption that the pp scenario has a $E^{-2}$ spectrum extended below $\sim 10$ TeV, which is not necessarily valid for all source types. For example, particles accelerated by fast-spinning newborn pulsars \citep{Blasi00, Fang:2012rx, Fang:2013vla} can have a spectrum  index  less than  2, and cosmic rays  accelerated in the AGN \citep{2012ApJ...749...63M, 2012ApJ...755..147D} can present a cutoff at low energies due to the confinement of the source environment.

In this paper, we investigate high-energy cosmic rays and neutrinos from clusters, by numerically propagating cosmic rays down to TeV in galaxy clusters with a wide range of  masses and redshifts.  We limit the uncertainties of our results with a more realistic modeling of the ICM gas and the turbulent cluster magnetic field than some of the previous works, and by adopting the mass accretion rates, cluster baryon fraction, and halo mass function as constrained by cosmological observations. 
 
We report the  integrated cosmic ray and neutrino spectra from the entire cluster population in two scenarios: 
\begin{enumerate} 
\item the {\it accretion shock scenario}: cosmic rays are  accelerated by the cluster accretion shocks and injected at the outskirts of clusters; 
 \item  the {\it central source scenario}: cosmic rays are  accelerated by sources at the centers of the clusters.
\end{enumerate}
Our results  demonstrate that
neutrinos from cluster accretion shocks could contribute $\lsim 20\%$ of the IceCube flux; however, if bright astrophysical sources inside the galaxy clusters can accelerate particles  
 with   an injection spectrum index  below 2, clusters could reproduce  both the spectrum and flux of IceCube neutrinos above $30$ TeV, while remaining consistent with the measurements of  high-energy cosmic rays and $\gamma$ rays. 

\section{Models}
\label{sec:models}

In this section we lay out the models used in our work. We first review the particle acceleration in clusters in Sec.~\ref{sec:acc}, then examine the diffusion and confinement of cosmic rays in the turbulent cluster magnetic field in Sec.~\ref{sec:diff}. We calculate the neutrino production in a  single cluster in Sec.~\ref{Sec:NvModel}, and discuss the formalism used to integrate the source contributions in Sec.~\ref{sec:integration}.

\subsection{Cosmic Ray Acceleration }\label{sec:acc}
 
Charged particles can get accelerated by large-scale shocks induced by cluster accretion activities, or (and) produced by the powerful sources harbored inside the clusters. Below we discuss the two scenarios accordingly. 

The spatial scale of the accretion shocks can be estimated by the  virial radius of the host cluster. For a cluster with mass $M=M_{15}\times 10^{15}\,M_\odot$ at redshift $z$, the size of the accretion shock is 
\bey
r_{\rm sh}\approx r_{\rm vir} &= &\left(\frac{3\,M}{4\pi\,\Delta_{\rm vir}(z)\rho_m(0)}\right)^{1/3} \\ \nonumber
&=& 2.6\, M_{15}^{1/3}\left(\frac{\Delta_{\rm vir}(0)}{\Delta_{\rm vir}(z)}\right)^{1/3}\,{\rm Mpc}
\eey
where $\triangle_{\mathrm {vir}} (z) \approx (18\pi^2 + 82 x - 39x^2)/(x+1)$ with $x =\Omega_M(z) - 1$ is the virial overdensity with respect to the mean matter density \citep{2003ApJ...584..702H}. For simplicity,  the cosmological term $\Delta_{\rm vir}(0)/\Delta_{\rm vir}(z)$  will be ignored  in the analytical estimations below (for reference, $\Delta_{\rm vir}(0)/\Delta_{\rm vir}(1)\approx 1.6$).

The typical   velocity of  accretion shocks should be comparable to the free-fall velocity \citep{Inoue07, Murase08}, which reads $v_{\rm sh}\approx \sqrt{{GM}/{r_{\rm sh}}} = 1300\, M_{15}^{1/3} \,\rm km\,s^{-1}$. 
The acceleration time in a strong shock can be estimated by  $t_{\rm acc}=20\,D_{\rm sh}/v_{\rm sh}^2$ \citep{ 1997MNRAS.286..257K,Inoue07}), assuming a quasi-parallel shock. Note this time can be much shorter for oblique shocks with more general magnetic field angles  \citep{1987ApJ...313..842J}. Strong accretion shocks are expected at the outskirts of clusters \citep{Vazza:2014usa}. 
Taking the Bohm limit, the diffusion coefficient of the cluster shock reads $D_{\rm sh}=r_L\,c/3$, where $r_L=E/ZeB$ is the Larmor radius of a cosmic ray particle with energy $E$ and charge $Z$ in a magnetic field $B$. In a shock with field strength $B=1\,B_{-6}\,\mu\rm G$, the acceleration time for a particle  to reach   $E_{\rm cr} = 10^{18}\,E_{\rm cr, 18}\,\rm eV$   would be 
$t_{\rm acc}  = 1.3\,E_{\rm cr, 18}\, M_{15}^{-2/3}\,B_{-6}^{-1}\,Z^{-1}\,\rm Gyr$.
On the other hand, the maximum time cosmic rays  stay in the shocks due to diffusive escape is $t_{\rm esc}\approx r_{\rm sh}^2/6\,D_{\rm sh}$ \citep{Inoue07}. $t_{\rm acc}=t_{\rm esc}$ thus sets the maximum energy that cosmic rays can be accelerated to:
\beq\label{eqn:Emax}
E_{\rm cr, max} = 2.8\times10^{18}\,M_{15}^{2/3}\,Z\,B_{-6}\,\rm eV.
\eeq

The energy budget of high-energy cosmic rays is determined by the accretion rate of the cluster. 
The mass accretion rate of halos can be described by a fitting function $\langle \dot{M}\rangle=42\,(M/10^{12}\,M_\odot)^{1.127}\,(1+1.17z)\,E(z)\,\rm M_\odot\,yr^{-1}$ \citep{McBride:2009ih}, with $E(z)=\sqrt{\Omega_m(1+z)^3+\Omega_\Lambda}$. Note that in this work we  take $\Omega_m=0.308$, $h=67.8$ \citep{2015arXiv150201589P} and assume a flat Universe. 

The kinetic energy of the accretion shocks is $L_{\rm acc}=f_b\,GM\dot{M}/r_{\rm sh}$, with $f_b=0.13\,(M/10^{14}M_\odot)^{0.16}$ being the average baryon fraction of galaxy clusters \citep{Gonzalez:2013awy}. If a fraction $f_{\rm cr}=1\%\,f_{\rm cr,-2}$ of this energy is converted into cosmic rays, the  luminosity of cosmic rays from  accretion shocks is then
\beq\label{eqn:L_cr}
L_{\rm cr} = 2.0\times10^{44}\,M_{15}^{1.95}\,f_{\rm cr,-2}\,\rm erg\,s^{-1}
\eeq
Notice  that the cosmic ray luminosity of a cluster is highly dependent on the cluster mass.

Particles accelerated in high-energy sources inside the clusters can have very different spectra and energetics depending on the source types. In general, $E^{-2}$ spectrum would be expected from diffusive   shock accelerations, but harder spectra are possible when the particle energy is converted from  the electromagnetic energy of the sources, for example, in pulsar magnetospheres \citep{Arons:2002yj}, or in reconnection processes \citep{2014PhRvL.113o5005G}. Besides, an interplay between the acceleration and   dissipation processes at the acceleration site could also lead to a hard injection spectrum \citep{2012ApJ...749...63M, 2012ApJ...755..147D}.  \cite{Murase:2015xka, Bechtol:2015uqb} demonstrated that a pp scenario with a $E^{-2}$ spectrum    normalized to the IceCube datapoint at  $\sim 10$ TeV  would  overproduce $\gamma$ rays above the flux of the non-blazar component of the isotropic $\gamma$-ray background. In this work we will instead consider the case of  an injection spectrum index equal or smaller than 2 but normalized  to the IceCube datapoint at 30 TeV (which is about 3 times lower than that at 10 TeV). Specifically, in Sec.~\ref{sec:results} we focus on a benchmark case of an injection spectrum $dN/dE\propto E^{-1.5}$ and  maximum energy $E_{\rm max} = 50\,\rm PeV$. We discuss the effects of different choices of $dN/dE$ and $E_{\rm max}$ in Sec.~\ref{sec:discussion}.

\subsection{Cosmic Ray Diffusion}\label{sec:diff}

Galaxy clusters are known as efficient containers of cosmic rays \citep{1996SSRv...75..279V, 0004-637X-487-2-529}. Depending on the particle's Larmor radius $r_L$ and the field's coherence length $l_c$, the propagation of cosmic rays in the cluster turbulent magnetic field can be divided into semi-diffusive ($r_L>l_c$)  and diffusive ($r_L\ll l_c$)  regimes \citep{Kotera08a}. 

When a particle's energy reaches $E_{\rm sd}=9.3\times10^{19}\,Z\,B_{-6}\,l_{c,-1}\,\rm eV$, the particle enters the semi-diffusive regime with a diffusion coefficient
\beq
D_{\rm cl} = \frac{1}{3}\,c\,r_L^2\,l_c^{-1}
\eeq
However, most particles have energy much lower than $E_{\rm sd}$, and thus stay in the diffusive regime. The diffusion coefficient in this regime  can be written as \citep{Brunetti:2014gsa}
\bey
D_{\rm cl} &=& \frac{1}{3}\,\left(\frac{ B}{\delta B}\right)^2\,c\,r_L^{2-w}\,l_c^{w-1} \\ \nonumber
&\approx&6.8\times10^{29}\,\left(\frac{l_c}{0.1{\,\rm Mpc}}\right)^{2/3}\,\left(\frac{E}{{\rm 1 GeV}}\right)^{1/3} \\ \nonumber
&\times& (B_{-6}Z)^{-1/3}\,\rm cm^2\,s^{-1}
\eey
where $w=5/3$ is the spectrum index for Kolmogorov  diffusion, $\delta B\sim B$ and $l_{0}=l_{0,-1}\,0.1\,\rm Mpc$ are the typical amplitude and  scale of magnetic field fluctuations in massive clusters \citep{2015ASSL..407..557B}. 

 The diffusion time of cosmic rays in the cluster magnetic field  can be estimated as 
\bey
t_{\rm diff}\approx \frac{r_{\rm vir}^2}{2D_{\rm cl}} =1.5\,M_{15}^{2/3}E_{18}^{-1/3}Z^{1/3}B_{-6}^{1/3}l_{c,-1}^{-2/3}\,\rm Gyr
\eey
Cosmic rays with diffusion time longer than the Hubble time $t_H\sim 14\,\rm Gyr$ would be completely confined by the cluster magnetic field, defining the lower limit of the energy of cosmic rays we expect to observe from the cluster as
\beq
E_{\rm cr, conf} = 1.4\times10^{15}\,M_{15}^2\,Z\,l_{0,-1}^{-2}\,B_{-6}\,\rm eV
\eeq

Particles with $E>E_{\rm cr, conf}$ would have a better chance to escape from the cluster magnetic field. To estimate this escape probability, we first write down the probability that a particle with diffusion coefficient $D$  reaches  radius $r$ after time $t$:
\beq
p(r, t) = \frac{3^{3/2}}{4\pi Dt}e^{-3r^2/4 Dt}
\eeq
assuming that the particle starts from $r=0$ and the field is homogenous. The probability that a particle can successfully escape from a cluster of size $r_{\rm vir}$ within the Hubble time is then  \citep{Kotera08a}
\beq
f_{\rm esc} \sim \int_{r_{\rm vir}}^\infty\,dr\,p(r, t_H)\,4\pi r^2
\eeq

Cosmic rays accelerated in the strong shocks in  the accretion region are expected to have a spectrum index $\alpha\sim 2$. The flux of high-energy cosmic rays from a cluster with mass $M=10^{15}\,M_\odot$ at redshift $z=0.1$ can be calculated as
\bey
E_{\rm CR}^2\Phi_{\rm CR}&=& E_{\rm CR}^2\frac{d\dot{N}}{dE}\frac{1}{4\pi d_L^2}\,f_{\rm esc}\,(1-f_\pi)\\ \nonumber
&=& 0.12\,\left(\frac{E}{1\,\rm PeV}\right)^{2-\alpha}\left(\frac{d_L}{d_L(z=0.1)}\right)^{-2} \\ \nonumber
&\times&  M_{15}^{1.95}f_{\rm CR,-2} f_{\rm esc}(1-f_\pi)\,\rm eV\,cm^{-2}s^{-1}sr^{-1}
\eey
where $d_L$ is the luminosity distance, and $f_\pi$ is the pion production rate that will be introduced in Sec.~\ref{Sec:NvModel}. Note that we have taken $\alpha\sim2.1$ in the  above estimation. In case of a much softer spectrum, an additional correction term $(\alpha-2)(\rm PeV/GeV)^{2-\alpha}$ originated from spectrum normalization should be included.

\begin{figure*}
\centering
\epsfig{file=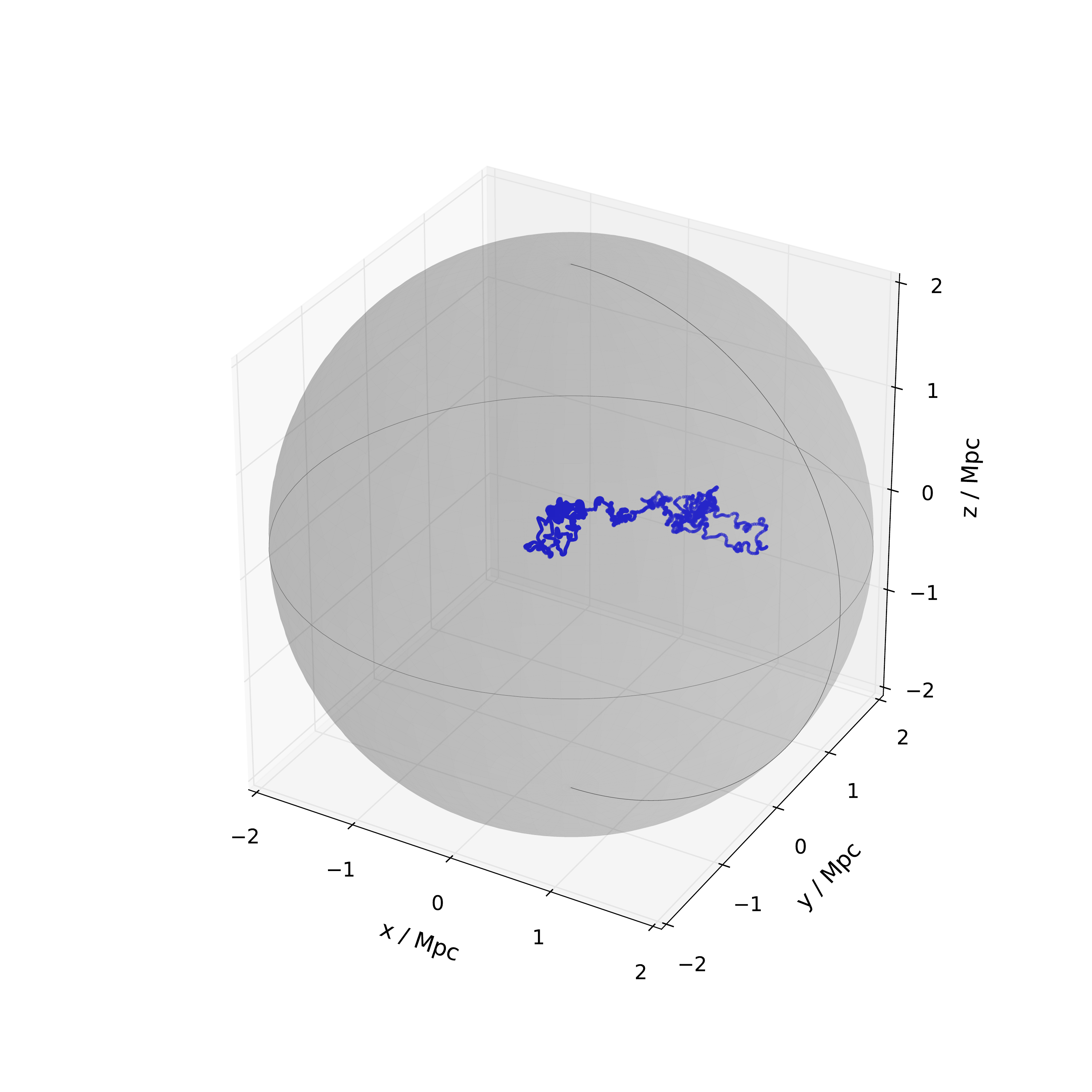,width=0.48\textwidth}  
\epsfig{file=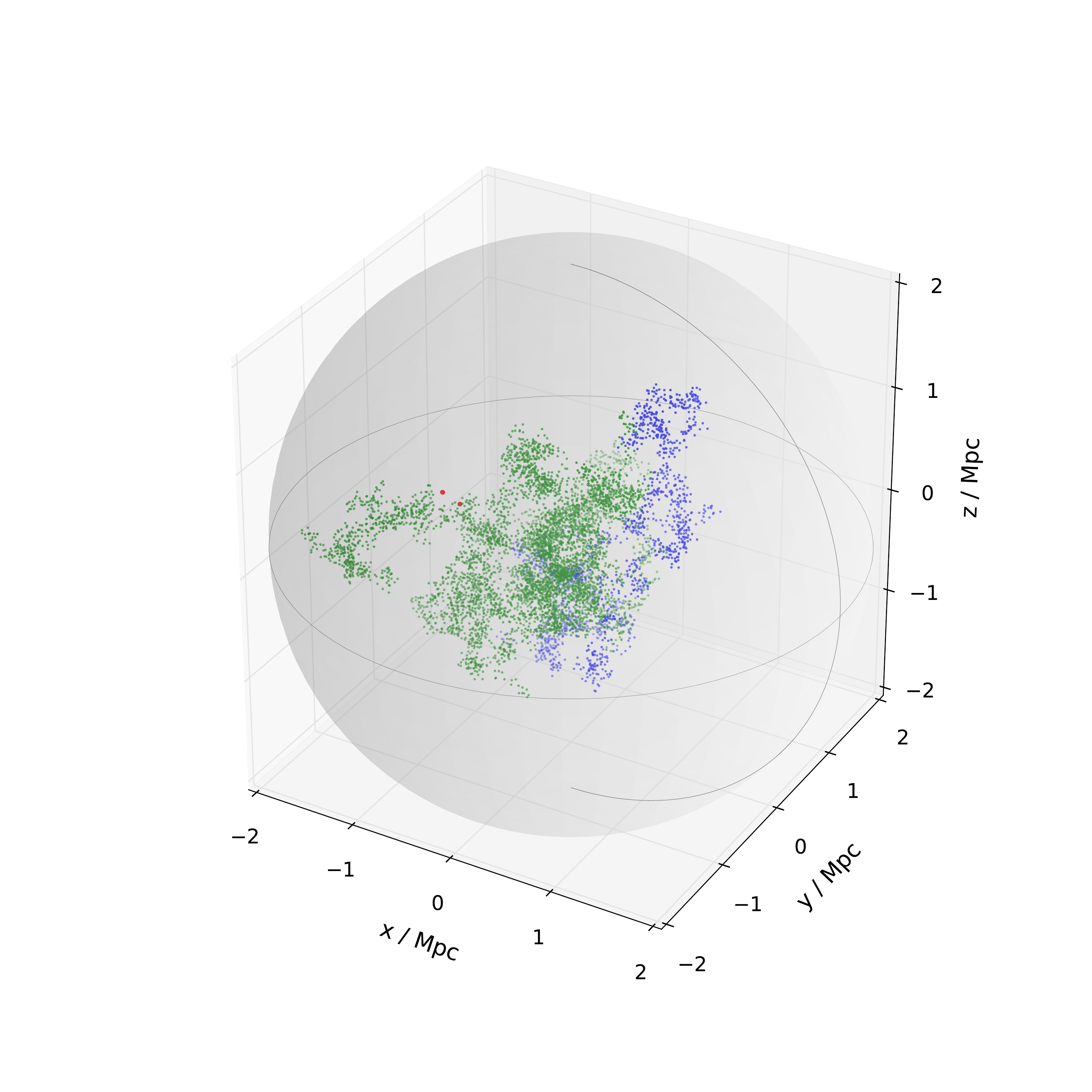,width=0.48\textwidth}  
\caption{\label{fig:traj} Trajectory of high-energy cosmic rays in the turbulent magnetic field of a model cluster with mass  $M=10^{15}\,M_\odot$ located at redshift $z=0.1$. 
The grey sphere represents a cluster. The virial radius of the simulated case is $2.6$ Mpc just outside the grey sphere.  The  baryon and the magnetic field  distribution in the ICM follows the ``$\beta$ model"  described by equation~\ref{eqn:n_ICM} and \ref{eqn:B_cluster}. The central magnetic field strength is  $B_0=10\,\mu$G. 
The left panel shows a proton with injection energy $E_p=10^{19}$ eV which has a Larmor radius comparable to the coherence length of the B field. The right panel shows the trajectories of a proton   with injection energy $E_p=10^{17}$ eV  and its interaction products. The starting point of the proton is  at the center of the cluster.
The trajectories of the primary proton (blue), the secondary protons (green), as well as the neutron and neutrino products (red)  are indicated by a series of points at each propagation step.}
\end{figure*}

\subsection{Neutrino Production}\label{Sec:NvModel}

Accelerated cosmic rays would propagate in the cluster magnetic filed and interact with both hadron  and photon ambience of the ICM.  The photon ambience of the ICM includes the Cosmic Microwave Background (CMB) and the intracluster infrared background, but both are subdominant when compared to  
 the hadronic background as shown in  \cite{Kotera09}.   We thus focus on the hadron background in this work. 

The density profile of the thermal gas in the ICM can be described by the classical ``$\beta$ model" \citep{1976A&A....49..137C}:
\beq\label{eqn:n_ICM}
n_{\rm ICM}(r) = n_{\rm ICM, 0}\,\left[1+\left(\frac{r}{r_c}\right)^2\right]^{-3\beta/2}
\eeq
where $\beta=0.8$ and $r_c\sim0.1\,r_{\rm vir}$ is the core radius \citep{Fujita03}. 
The density profile is normalized by  
\beq \label{eqn:n_ICM_0}
f_{\rm b}  M = \int \rho_{\rm ICM}(r) 4\pi r^2dr
\eeq
The mass density is related to the number density through $\rho_{\rm ICM}=n_{\rm ICM}\,\mu m_p$, with $\mu\approx0.61$ being the mean molecular weight.

The magnetic field distribution should trace the baryon distribution in the cluster through the flux conservation $B\propto n_{\rm th}^{2/3}$ \citep{Brunetti04}. Specifically, in cluster-sized halos, which we define here to consist of halos greater than $10^{12}$~M$_\odot$,   the magnetic field  as a function of the radial distance from the center of the cluster is given by 
\label{subsec:bfield}
\begin{equation}\label{eqn:B_cluster}
B(M,r) = B_0 \left(\frac{M}{M_{0}}\right)^\lambda\left[1 + \left(\frac{r}{r_{\mathrm{c}}}\right)^2\right]^{-\beta}
\end{equation}
where $B_0$ is the central magnetic field at a mass M$_0$ = 10$^{14}$M$_\odot$, and  $\lambda$ controls the dependence of $B_0$ on the mass of the cluster in consideration. In our default model, we set $B_0=3\,\mu$G and $\lambda= 0.0$ \citep{1999APh....11...73V,2012SSRv..166..215B}.  Following the distribution of $n_{\rm ICM}$, we define the core radius of the magnetic field model to be equivalent to r$_c$~=~0.1~r$_{\rm vir}$ and the falloff index $\beta=0.8$. The magnetic fields in galaxy-sized halos are instead dominated by the dense baryonic cores, and thus have a different structure. We thus use a two part magnetic field model which adopts the larger of the two following magnetic fields as in \cite{Fang:2014joa}:

\setlength{\thinmuskip}{2mu}
\begin{equation}\label{eqn:B_galaxy}
B(r) =\max\left( B_1 \,e^{-r/R_1},  B_2\,e^{-r/R_2}\right)
\end{equation}
where B$_1$ = 7.6 $\mu$G and R$_1$ = 0.025~R$_{\mathrm{vir}}$ and B$_2$~=~35~$\mu$G with R$_2$ = 0.008R$_{\mathrm{vir}}$. However we note that due to the sharp cutoff around $R_1$ and $R_2$, the magnetic field at large radii, where accretion shocks most likely occur, are too low to accelerate to high energies. The contribution from shocks in low mass halos is rather negligible in our scenario.

For one crossing of the cluster, a particle would expect to confront a column density of
\beq
\langle y_{\rm ICM}\rangle \sim \int_0^{r_{\rm vir}} n_{\rm ICM}(r)\,dr = 1.2\times 10^{22}M_{15}^{0.5}\,{\rm cm^{-2}} 
\eeq
This corresponds to a pion production rate:
\beq 
f_{\rm pp}=\langle y_{\rm ICM}\rangle \sigma_{\rm pp} \approx 6\times 10^{-3}\, M_{15}^{0.5}
\eeq
where $\sigma_{\rm pp} \sim 5\times 10^{-26}\,\rm cm^{-2}$ is the effective proton-proton interaction cross section. Note that particles injected at the cluster center would have a much greater chance to be confined by the cluster magnetic fields which leads to larger $f_{\rm pp}$. In contrast, particles injected near the boundary    may not even meet an ICM baryon before escaping from the cluster.  Since the distribution of baryons in the cluster is not uniform, and different confinement time of particles in the cluster magnetic field would lead to different  $f_{\rm pp}$,    in simulations this rate is more precisely calculated by tracking particle propagation in the cluster magnetic field (see Sec~\ref{sec:numericalSetup}).

\subsection{Integrated Neutrino Flux}
\label{sec:integration}

We expect a diffusive neutrino flux integrated from the entire clusters  population. 
The differential number density,   $dn/dM$, of clusters with mass M at redshift z is given by the  halo mass function:
\begin{equation}
\label{eq:dndM}
\frac{dn}{dM} (M, z)= f(\sigma) \frac{\rho_m}{M}\frac{d \ln~\sigma^{-1}}{dM}
\end{equation} 
where $\rho_m$ is the mean density of the universe at the epoch of analysis, $\rho_m(z) =\rho_m(0)\,(1+z)^3$, and $\sigma(M, z)$   is the rms variance of the linear density field smoothed on a top-hat window function $R=(3M/4\pi\rho_m)^{1/3}$. Specifically,
$\sigma^2 = \int  dk\, P^{\mathrm {lin}}(k) \tilde W(kR)\,k^2$, where $\tilde W(kR)$  is the Fourier transform of the real-space top-hat window function of radius R \citep{2008ApJ...688..709T}. For $f(\sigma)$ we adopt the mass function multiplicity described by  \citet{sheth_tormen_mass_function}.

The integrated neutrino flux can be calculated as 
\bey
E_\nu^2\Phi_\nu (E_\nu) &=&   \int dM \frac{dn}{dM}\frac{(1+z)^2L_\nu(M,z)}{4\pi d_L^2}\,\rm \frac{dV}{d\Omega} \\ \nonumber
&=&\frac{1}{4\pi} \int_{z_{\rm min}}^{z_{\rm max}}\frac{c\,dz}{H_0 E(z)}\\ \nonumber
&\times&\int_{M_{\rm min}}^{M_{\rm max}}\,dM\,\frac{dn}{dM}E_\nu^2\frac{d\dot{N}}{dE}((1+z)E_\nu,M,z)
\eey
where $dV={c\,dz}/({H_0E(z)})(1+z)^2d_A^2d\Omega$ is the comoving volume and $d_A=d_L(1+z)^{-2}$ is the angular diameter distance. The lower and upper limits of the integration are taken to be $M_{\rm min}=10^{12}\,M_\odot$ and $M_{\rm max}=10^{16}\,M_\odot$, since galaxies with  $M<M_{\rm min}$ barely contribute  to the energy window we are interested in, while   clusters with $M>M_{\rm max}$ have too low a number density to significantly contribute to the total flux. The redshift integration goes from $z_{\rm min}=0.01$, where the closest galaxy clusters are located, up to $z_{\rm max}=5$. As we will demonstrate in Sec.~\ref{sec:spectra}, the flux from the most   distant clusters is  dominated by that from the closer ones, so our results are not sensitive to the choice of $z_{\rm max}$ as long as $z_{\rm max}\gg 1$.

\section{Results}\label{sec:results}

\subsection{Numerical Setup}\label{sec:numericalSetup}

The interactions between high-energy proton and ICM baryons were simulated by Monte Carlo as  in \cite{FKO12}. The cross sections and  products of  pp interactions were calculated based on the hadronic interaction model EPOS \citep{PhysRevC.74.044902}. In addition, we implemented the public UHECR propagation code CRPropa3 \citep{2013APh....42...41K} to study the semi-diffusive propagation of high-energy cosmic rays in  cluster magnetic fields. 

In our simulation, the baryon distribution in the ICM is set to follow equation~\ref{eqn:n_ICM}. The field is generated to follow the Kolmogorov-type turbulence  with spectrum index $w=5/3$ and have random directions with a coherence length $l_c\sim 0.03\,r_{\rm vir}$ \citep{2015ASSL..407..557B}. The strength of the  magnetic field of the ICM is  normalized by equations~\ref{eqn:B_cluster} and \ref{eqn:B_galaxy}. (We will discuss the impact of a stronger or weaker $B$  on our results in Sec.~\ref{sec:discussion}.)  Therefore in our model the cluster environment is fully determined by two parameters: the cluster mass $M$ and the redshift $z$. 

In the accretion shock scenario,  we assume that a stationary shock is located at $d_{\rm sh}=\lambda_{\rm sh}R_{\rm ta}$ \citep{1998APh.....9..227C}, where $\lambda_{\rm sh}\approx0.347$, and $R_{\rm ta}\approx 2\,R_{\rm vir}$ is the turn-around radius. The maximum energy of the injected cosmic rays is calculated by equation~\ref{eqn:Emax}  taking  $B=B(d_{\rm sh})$. Note that adopting the magnetic field strength at the cluster boundary is a conservative assumption as the shocked region in a cluster is not limited to the outer skirts.  The injection of cosmic rays is assumed to follow  a spectrum  $dN/dE\propto E^{-2}$  normalized by equation~\ref{eqn:L_cr}. The  energy fraction $f_{\rm cr}$ is left as a free parameter to be determined by  observation.
 
In the central source scenario, cosmic rays are injected in isotropic directions from the center of the host cluster, with a universal maximum energy $E_{\rm max} = 50\,\rm PeV$ and spectrum $dN/dE\propto E^{-1.5}$. The effect of  different choices of $E_{\rm max}$ and $dN/dE$ to our results will be discussed in Sec. \ref{sec:discussion}.  Considering that the total luminosity of high-energy sources inside the cluster should scale to the kinetic power of the cluster, we still refer to equation~\ref{eqn:L_cr} as  the  energy budget of injected cosmic rays. 
 
Finally, the diffusive propagation of each particle is tracked until either it successfully leaves the cluster, or   its total propagation time exceeds the Hubble time. 
In the accretion shock scenario, particles are injected isotropically from $d_{\rm sh}\sim 0.7\,R_{\rm vir}$. Both magnetic field and number density of the ICM are low at the outskirt region, hence some fraction of the particles can leave the cluster within a relatively short time and nearly zero interaction, whereas the others will be trapped toward the inner side of the cluster.  
In the central source scenario, particles injected from the cluster center confront the densest region of the ICM, and could be easily confined over the Hubble time.  Meanwhile secondary and higher-order protons resulted from the interactions can  lead to further productions of neutrinos (an example of the presence of such secondary particles are shown by  the green points in the right panel of Fig~\ref{fig:traj}). 
The difference in the injection locations lead to quite different behaviors of particles in the two scenarios. 
The 3-d propagation and particle tracking in our simulation are needed to describe the transportation and interaction of particles of all orders in   the  non-uniform    environment, especially when the system is  asymmetric as in the accretion shock scenario.

\subsection{Cosmic Ray Trajectories}
In this section we demonstrate the propagation of cosmic rays in the turbulent  magnetic field of a model cluster with mass $M=10^{15}\,M_\odot$ at redshift $z=0.1$. The  central magnetic field strength  is  set to be $B_0=10\,\mu$G. The field strength at the outer skirts is then $B_{\rm vir} \approx 0.6\,\mu$G. The number density of the ICM gas at the cluster center is $n_0=1.1\times 10^{-2}\,\rm cm^{-3}$, as calculated  by equation~\ref{eqn:n_ICM_0}.

In the left panel of Figure~\ref{fig:traj}, we show the trajectory of a proton with  injection energy $E_p=10^{19}\,\rm eV$.  Note that the cluster radius is 2.7 Mpc although it looks smaller due to the projection of the 3D sphere in the plots.
The Larmor radius of the particle  is $r_L=0.01\,E_{19}B_{-6}^{-1}\,\rm Mpc$,  comparable to the coherence length of the field which is set to be $l_c = 0.07 \,\rm Mpc$.  To study such a semi-diffusive propagation, we utilize the  Cash-Karp propagation method of CRPropa   to track the particle's entire trajectory. Each step of the propagation is indicated as a blue point in the plot, and the step size is determined dominantly by the Larmor radius,  because  the mean free path of the charged particle is much larger than $r_L$. As expected, the particle travels almost rectilinearly. In this realization, the particle escaped the cluster without interaction with the ICM gas, and had a total trajectory length of 46 Mpc. 

In the right panel of Figure~\ref{fig:traj}, we show the trajectory of  a proton with  injection energy  $E_p=10^{17}$ eV. The Larmor radius of the particle is now $\sim$ 700 times less than the coherence length of the field, placing it in the fully diffusive regime. The computing cost for tracking all the steps of the particle would be enormous. Instead, we use the approximation algorithm demonstrated in \cite{Kotera08a}.   Specifically, taking advantage of the fact that a particle would lose its initial direction after traveling for a  distance much longer than the Larmor radius, the particle can be simulated as undergoing a random walk with step size $l_c$, but has each numerical step  correspond to an actual trajectory length $l_c^2/2D$, with $D$  the diffusion coefficient  at the particle's current location.  The points in the right panel of Figure~\ref{fig:traj}  hence   describe such a random walk with a uniform step size $l_c$. Note however these steps represent different  actual trajectory lengths  due to the distinctive $D$  at different parts of the cluster. The blue points correspond to the trajectory of the primary proton, the green  ones indicate that of the secondary protons   stemming from the interaction  with the ICM, and lastly   the red ones   mark the neutron and neutrino products, including both electron and muon neutrinos.  The primary proton was never able to leave the cluster, though the neutrons and some of the secondary protons  succeeded.  We conclude this section by  noting that the trajectories are random and could be very different from realization to realization. 

\subsection{Cosmic Ray and Neutrino Spectra}\label{sec:spectra}

\begin{figure}
\centering
\epsfig{file=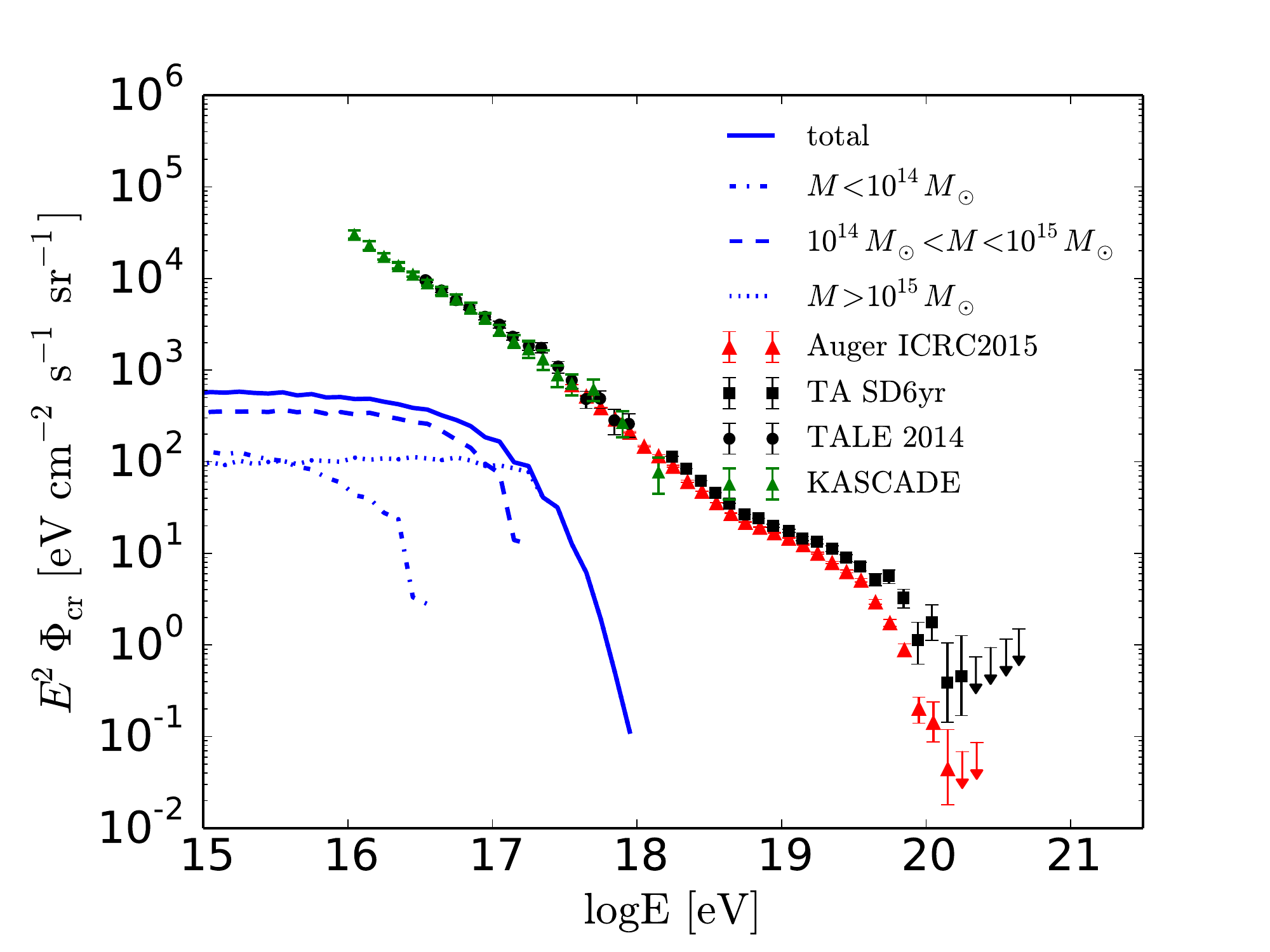,width=0.45\textwidth}  
\caption{\label{fig:PhiCR}
 Cumulative cosmic ray flux from the accretion shocks of galaxy clusters in comparison with observational data from KASCADE \citep{::2013dga}, Auger \citep{Aab:2015bza} and Telescope Array \citep{Fukushima:2015bza}. Cosmic ray protons are injected with a spectrum $dN/dE \propto E^{-2}$ and a $f_{\rm cr} = 2\%$ conversion rate from accretion energy to cosmic rays.    The solid line indicates  the total  flux  of cosmic rays that escape the cluster magnetic field within the Hubble time. The contribution is decomposed into three mass groups: $M<10^{14}\,M_\odot$ (dash dotted), $10^{14}\,M_\odot<M<10^{15}\,M_\odot$ (dashed), $M>10^{15}\,M_\odot$ (dotted).  The central magnetic field strength is assumed to be $3\,\mu$G for all  cluster halos in this scenario. }
\end{figure}

\begin{figure}
\centering
\epsfig{file=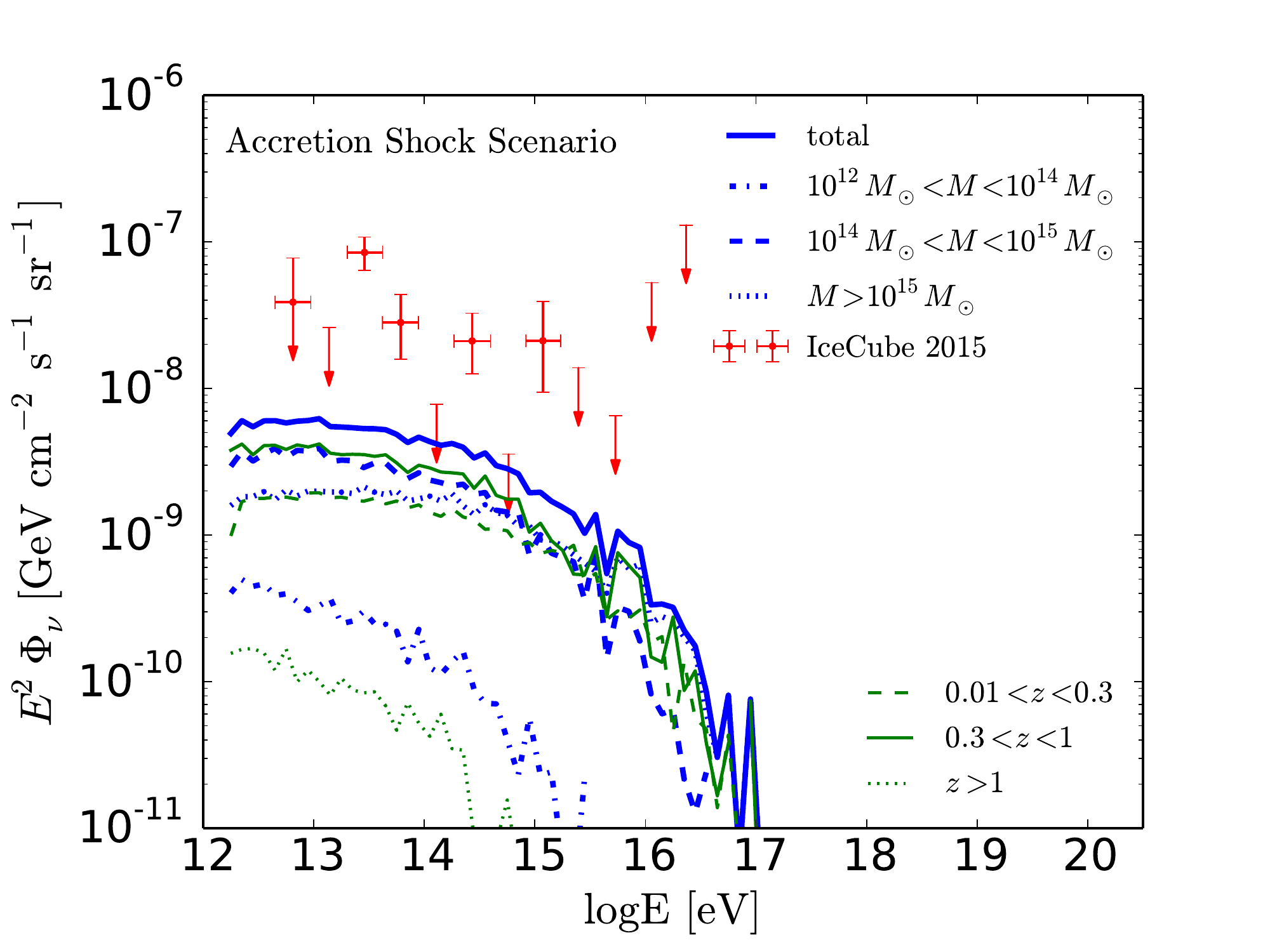,width=0.45\textwidth}  
\epsfig{file=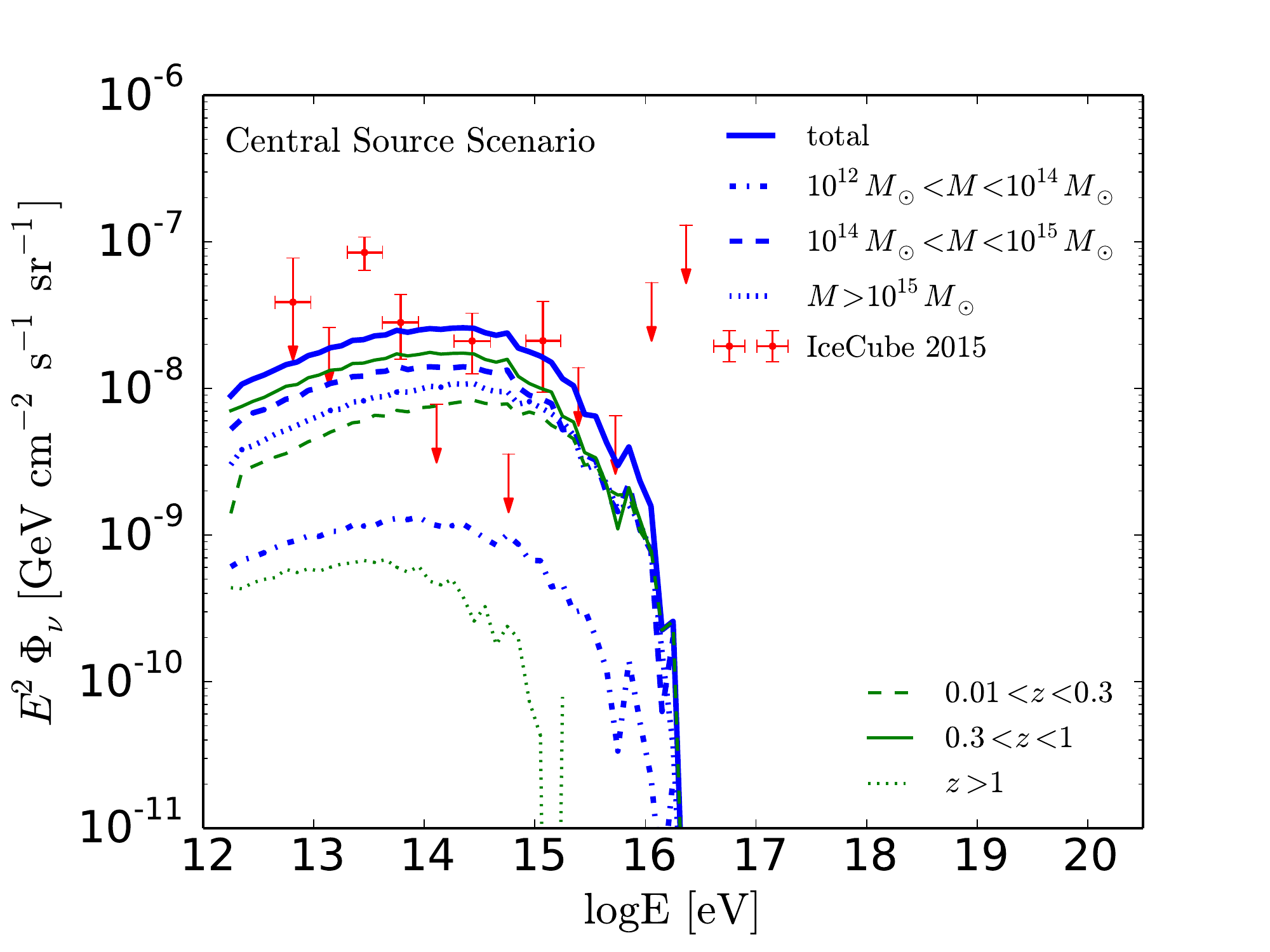,width=0.45\textwidth}  
\caption{\label{fig:PhiNeu} Cumulative neutrino flux from the  galaxy clusters compared to the IceCube observations \citep{2015PhRvD..91b2001A}. Top: accretion shock scenario: as in Figure~\ref{fig:PhiCR}, the cosmic ray injection follows $dN/dE\propto E^{-2}$, $f_{\rm cr} =  2\%$, and a maximum energy following equation~\ref{eqn:Emax}, determined by particle acceleration and escape in the accretion shocks. Bottom: central source scenario: the cosmic ray injection  follows $dN/dE\propto E^{-1.5}$, $f_{\rm cr} = 0.5\%$, and $E_{\rm max}=5\times10^{16}$ eV. 
The blue thin lines decompose the contribution into  three mass groups:  $M < 10^{14}\,M_\odot$ (dash dotted) and $10^{14}\,M_\odot < M < 10^{15}\,M_\odot$ (dashed) and $M > 10^{15}\,M_\odot$ (dotted).  The green thin lines   instead decompose the flux  into three redshift bins: $0.01<z<0.3$ (dashed), $0.3<z<1$ (solid) and $z>1$(dotted).}
\end{figure}

The integrated cosmic ray flux from the cluster population  in the accretion shock scenario is shown in Figure~\ref{fig:PhiCR}. In all clusters in this scenario, cosmic rays are injected as protons following a power-law spectrum $dN/dE\propto E^{-2}$. The conversion rate from the kinetic energy of the cluster shocks to the cosmic rays is set to be $f_{\rm cr} = 2\%$ to fit the observed neutrino flux. The turbulent magnetic fields in clusters have a central strength $B_0=3\,\mu$G, and the baryon density of the ICM is scaled to the cluster mass by $n_0\propto f_b\propto M^{0.16}$. 
 The solid line indicates the flux of the cosmic rays that could successfully leave the clusters and  be observed at the earth. The overall spectrum follows $E^{-2}$ as the injection spectrum  and is cut off around $10^{17.5}$ eV. The  contribution to cosmic ray flux from  clusters  is subdominant in all energies comparing to observation. The integrated cosmic ray flux in the central source scenario has a similar behavior as in the accretion shock scenario, except that we assume the flux follows a $E^{-1.5}$ injection spectrum. 

We further decompose the  flux into the contribution   from three mass groups: $M<10^{14}\,M_\odot$, $10^{14}\,M_\odot<M<10^{15}\,M_\odot$, $M>10^{15}\,M_\odot$, shown in dash dotted, dashed, and dotted lines correspondingly. As the cosmic ray luminosity significantly depends on the cluster mass through $L_{\rm cr}\propto M^{1.95}$, clusters below $10^{14}\,M_\odot$ are not as luminous as the massive clusters. In addition, since $E_{\rm max}\propto M^{2/3}$,  particles from these clusters cannot reach the highest energies.   However, since the strength and scale of magnetic fields in these clusters are also smaller,  particles have a better chance to leave the source. The combined impact of these factors leads to a contribution of this mass group to $\sim 1/3$ of the total cosmic ray flux around PeV, and  $\le 10\%$ above $10^{17}\,\rm eV$. The $10^{14}\,M_\odot<M<10^{15}\,M_\odot$ group is found to make the largest contribution to the flux up to the ankle around $10^{18}\,\rm eV$.  

Above the ankle, clusters with mass $M >10^{15}\,M_\odot$ are needed to produce UHE protons. However, due to the exponential cutoff in the halo mass function for such extremely massive clusters,  a significant  production of UHECRs from clusters is not found given a pure proton injection. The main contribution of this proton component  is relevant to explain the second knee observed around $10^{17}$ eV \citep{::2013dga}. Heavier nuclei, if injected at the shocks, may reach ultra-high energies, but their neutrino production would be less significant. 

As displayed in Figure~\ref{fig:traj}, within the same cluster, cosmic rays with lower energy have a lower chance to leave the system due to their smaller Larmor radius. Consequently, the spectrum of the escaped cosmic rays is slightly harder than that at  injection.  
 
The cumulative neutrino flux from  clusters is shown in Figure~\ref{fig:PhiNeu}.  The top panel corresponds to the accretion shock scenario. The injection of cosmic rays and the cluster environment are set up as in Figure~\ref{fig:PhiCR}. The total neutrino flux is indicated by the solid blue line. With $f_{\rm cr}=2\,\%$,  clusters  contribute  a neutrino flux less than $20\%$ of the IceCube detections. Unlike the cosmic ray spectrum which mainly scales with the escape probability $f_{\rm esc}$, the neutrino spectrum is rather sensitive to the pion production rate $f_\pi$.  The spectral index is found to be about 2 around TeV, 2.4 at PeV and cut off above 10 PeV. The spectrum softening at PeV is a combined effect of  diffusion (lower $f_\pi$ at higher energies) and cluster mass function (higher flux but lower rate for massive clusters). 
  
The bottom panel of Figure~\ref{fig:PhiNeu} shows the integrated neutrino flux in the central source scenario. The spectrum below $\sim$100 TeV is mainly determined by the injection and thus follows $\sim E^{-1.5}$. Interestingly, at higher energies the softening effect due to the shorter cosmic ray diffusion time and the lower population of massive clusters dominates over the hard injection.  As a result, with $f_{\rm cr} = 0.5\%$ this scenario could reproduce the spectrum and flux of the IceCube measurement above $\sim 30$ TeV.

As in  Figure~\ref{fig:PhiCR}, we also decompose the neutrino contribution  into three mass groups: $M<10^{14}\,M_\odot$ (dash dotted), $10^{14}\,M_\odot < M < 10^{15}\,M_\odot$ (dashed) and $M >10^{15}\,M_\odot$ (dotted), all indicated by  blue lines.  The significance of the three mass groups' contribution to the neutrino fluxes follows a similar order as that of cosmic rays, except that without the $f_{\rm esc}$ factor impacting the neutrino production, the $M<10^{14}\,M_\odot$ group now contributes less than $5\,\%$ of the total flux. 
The thin green lines in Figure~\ref{fig:PhiCR}   decompose the contribution into three redshift bins:   $0.01<z<0.3$ (dashed), $0.3<z<1$ (solid) and $z>1$(dotted). We find that  due to the intense source distribution in the region $0.3<z<1$, this group makes the largest contribution to neutrino production. Due to the distance and the rare rate of massive clusters at high redshift,  clusters at $z>1$ barely contribute.

\section{Discussion and Conclusion}\label{sec:discussion}

Unlike other astrophysical sources, galaxy clusters offer a unique environment for TeV-PeV neutrino production through efficient acceleration and confinement of high-energy cosmic rays. By propagating particles in three-dimensional turbulent magnetic fields and recording their interactions with the ICM gas, we find that the integrated neutrino flux from the cluster accretion shocks could account for $\le 20\%$ of the IceCube detections, while neutrinos produced by the interaction of cosmic rays from powerful sources hosted by clusters could explain both the flux and spectrum of the IceCube data above 30 TeV, if the injection spectrum is  harder than 2. The lack of neutrino clustering around known sources also fits well with the cluster production model. 
In addition,  the high-energy cosmic rays that succeed in escaping from the clusters contribute less than 20\% of the observed cosmic ray flux around and below the ankle, and may explain the second knee feature when galactic and ultrahigh energy cosmic ray accelerators contributions are added to the clusters' contribution.

Our results demonstrate that when taking into account the effect of cosmic ray confinement and the mass dependence of the cluster number density,  the integrated neutrino spectrum would conserve the injection spectrum around TeV - 100 TeV, but become much steeper above PeV. If high-energy sources harbored in the galaxy clusters could produce cosmic rays with spectrum   harder than $E^{-2}$, the tension between the neutrino-associated GeV  $\gamma$ rays and   the Fermi measurement of isotropic  diffusive $\gamma$-ray background \citep{0004-637X-799-1-86,Murase:2015xka} can be alliviated.   Also notice that in the cluster scenario, the dominant contribution  comes from sources beyond $z\sim0.3$ (as suggested by Fig.~\ref{fig:PhiNeu}). Thus,  $\gamma$ rays with $E\gsim$0.1 TeV  are  attenuated by the time they arrive on Earth, through interactions with photons of the extragalactic background light (EBL) and the CMB that have a non-trivial optical depth, $\tau_{\gamma\gamma}(z\sim0.3)> 1$ \citep{2006ApJ...648..774S}.

The search for the first $\gamma$ rays from galaxy clusters is still ongoing. No significant spatially extended $\gamma$-ray emission from the nearby galaxy clusters was found in four years of Fermi-LAT data, establishing limits on the cosmic ray to thermal pressure ratio, $X_{\rm CR}$, to be below 1.4\% \citep{2014ApJ...787...18A}. The parameter $f_{\rm CR}$ of our model can be translated to  $X_{\rm CR} $ by $X_{\rm CR} = P_{\rm CR} / P_{\rm th}\sim 0.6\%\,f_{\rm CR,-2}\, (\dot{M}\,t_{\rm cr, conf}/M)$, where $P_{\rm CR}\approx (1/3)\,f_{\rm CR}(G\dot{M}\rho_{\rm ICM}\,t_{\rm cr,conf}/r_{\rm vir})$ is the cosmic ray pressure after an accumulation of particles for $t_{\rm cr, conf}$, and $P_{\rm th } =   n_{\rm ICM} (GM\mu m_p/ 2\,r_{\rm vir})$ is the thermal pressure. The  $f_{\rm CR} <2\%$  suggested by our model is thus consistent with the constraint from $\gamma$-ray searches. However, the Fermi limit would constrain cluster scenarios with $f_{\rm CR}>2\%$,  corresponding to an injection index $\alpha>2.1$.

In the central source scenario (bottom panel of Fig.~\ref{fig:PhiNeu}) we showed a benchmark case with an injection spectrum index $\alpha = 1.5$ and maximum cosmic ray energy $E_{\rm max} = 50\,\rm PeV$. The results are solid under  moderate  adjustments on either $\alpha$ or $E_{\rm max}$ (including  adding a mass dependence).   However, $\alpha$ cannot be as large as $\sim$ 2,  otherwise the $\gamma$-ray counterpart will clearly exceed the Fermi measurements of the isotropic $\gamma$-ray background. Furthermore, if $E_{\rm max}\ll 50 \,\rm PeV$,  the injected cosmic rays wouldn't be energetic enough to produce the PeV neutrinos. On the other hand, if $E_{\rm max}\gg 50$ PeV in all clusters, the resulted neutrino spectrum could overshoot the null bins at  5 - 10 PeV. 

In our simulations we assumed a central magnetic field strength of $3\,\mu$G. In case of a smaller central magnetic field strength $B_0$, highest energy cosmic rays would have a higher chance to leave the system due to the weaker magnetic field and less interaction materials in the environment. Lower energy cosmic rays would be less impacted, since they are confined to a relatively small volume.  Hence the overall neutrino spectrum would be expected to have a lower flux but a softer spectrum. Conversely, stronger $B_0$ would lead to a higher neutrino flux due to the intenser confinement. 

Neither of our scenarios could account for the IceCube data below 30 TeV. Although majority of the neutrinos are expected to come from extragalactic sources, Galactic sources could  potentially   contribute some fraction of the flux  \citep{Ahlers:2015moa}, especially below $\sim 200 \,\rm TeV$ \citep{MuraseReview}. It is also possible that other types of sources could contribute to this energy range.

Our conclusion on  the accretion shock scenario is  consistent with that from \cite{2015A&A...578A..32Z}, though the two works have distinctive approaches. While \cite{2015A&A...578A..32Z}  derived the neutrino luminosity from a scaling  between  radio and gamma-ray luminosities, we directly simulated the particle propagation to obtain the  neutrino spectrum. \cite{2015A&A...578A..32Z} did not assume any cosmic ray spectral steepening due to the escape of high-energy cosmic rays, whereas in our work this cutoff was shown as a natural result from the energy-dependent transportation, and crucial for the survival of the central source scenario. On the other hand,  \cite{2014MNRAS.438..124Z, 2015A&A...578A..32Z}  indicated that  the neutrino contribution from clusters could be further limited  due to the fact that not all   clusters are expected to produce  hadronic emissions. The same constraint could also apply to our  results in the accretion shock scenario.  
 
Although consistent with current $\gamma$-ray and radio detection limits, the cluster scenario could be robustly tested by the growing statistics of IceCube as well as $\gamma$-ray searches of galaxy clusters in the near future. If  the cluster contribution to the diffusive $\gamma$-ray background and the cosmic ray to thermal pressure ratio can be further   constrained by the $\gamma$-ray observations, a significant contribution to IceCube neutrinos from galaxy clusters would be ruled out. On the other hand, 
as our model predicts a dominant neutrino flux from the most massive clusters above $\sim$5 PeV, future detection of strong anisotropy in this energy regime would provide a firm support to the cluster scenario.

\acknowledgments
We thank Cole Miller, Kohta Murase for helpful discussions. 
KF  acknowledges the support of a Joint Space-Science Institute prize postdoctoral fellowship. 
AO acknowledge financial support from the NSF grant NSF PHY-1412261 and  the NASA grant 11-APRA-0066 at  the University of Chicago, and the grant NSF PHY-1125897 at the Kavli Institute for Cosmological Physics. This work made use of computing resources and support provided by the Research Computing Center at the University of Chicago.

\bibliography{FO15}

\end{document}